\documentclass[pra,preprint,tightenlines]{revtex4}
\usepackage{amsmath,amssymb,pictex,color}
\usepackage{tikz}
\usepackage[normalem]{ulem}
\usepackage{braket}
\usetikzlibrary{calc}
\newcommand{\beq}{\begin{equation}}
\newcommand{\eeq}{\end{equation}}

\begin{document}
\title{The Equivalence Principle at Work in Radiation from 
Unaccelerated Atoms and Mirrors}

\author{S. A. Fulling} 
\email{fulling@math.tamu.edu}
 \homepage{http://www.math.tamu.edu/~fulling}
\affiliation{Department of Mathematics,  Texas A\&M 
University, College Station, TX 77843-3368, USA}
 \affiliation{Institute for Quantum Science and 
Engineering and Department of 
Physics and Astronomy, Texas A\&M 
University,   College Station, TX 77843-4242, USA}
\author{J. H. Wilson}
\email{jwilson@caltech.edu}
\affiliation{ Institute of Quantum Information and Matter and 
Department of Physics, California Institute of Technology, 
Pasadena, CA 91125 USA}

\date{\today}

\begin{abstract}
The equivalence principle is a perennial subject of controversy, 
especially in connection with radiation by a uniformly 
accelerated classical charge, or a freely falling charge observed 
by a supported detector.  Recently, related issues have been 
raised in connection with the Unruh radiation associated with 
accelerated detectors (including two-level atoms and resonant 
cavities).  A third type of system, very easy to analyze because 
of conformal invariance, is a two-dimensional scalar field 
interacting with perfectly reflecting boundaries (mirrors).
After reviewing the issues for atoms and cavities, we investigate 
a stationary mirror from the point of view of  an accelerated 
detector in ``Rindler space''\negthinspace.
In keeping with the conclusions of earlier authors about the 
electromagnetic problem, we find that  a radiative 
effect is indeed observed; from an inertial point of view, the 
process arises from a 
collision of the negative vacuum energy of Rindler space with the 
mirror.
There is a qualitative symmetry under interchange of accelerated 
and inertial subsystems (a vindication of 
the equivalence principle), but it hinges on the accelerated detector's 
being initially in its own ``Rindler vacuum''\negthinspace.
This observation is consistent with  the recent work on the Unruh 
problem.
\end{abstract}

\maketitle

\section{Introduction} \label{sec:intro}

Wolfgang Schleich is a master at combining quantum theory 
(particle or field) with relativity (special or general).
It is a privilege to offer the present work in his honor.

In fact, the immediate impetus to this work came from a project 
in which Wolfgang is involved \cite{SFLPSS}. It concluded that  
atoms falling from outside through a cavity into a black hole 
 emit acceleration radiation which to a distant observer looks 
much like Hawking black-hole radiation \cite{haw}. 
 The derivation is a 
straightforward variant (following \cite{SKBFC,BKCFZS}) of the 
Unruh--Wald \cite{U76,UW} 
analysis of the behavior of a uniformly accelerated detector.
According to general relativity, however, the atom of 
\cite{SFLPSS} is in free fall (not accelerated), and the cavity 
is accelerated (supported against the gravitational field of the 
black hole).  The fact of \emph{relative} acceleration is 
critical to the result.  A follow-up paper \cite{SBFP} 
perturbatively confirms this picture by 
showing in flat space-time that uniformly accelerated motion of a 
mirror can yield excitation of a two-level atom 
moving at constant velocity,
with simultaneous emission of a real photon; in this calculation 
the photon mode considered is initially empty in the Rindler 
sense (i.e., with respect to field quantization based on the 
timelike Killing vector field of Lorentz boosts along the 
hyperbolic mirror trajectory).

The term ``principle of equivalence'' has been used to mean many 
different things.  The version we have in mind here is that 
``uniform gravitational fields are equivalent to frames that 
accelerate uniformly relative to inertial [free-falling] 
frames'' \cite[p.~115]{schutz}.  Precisely what this means can be 
subtle.  Physicists have argued for decades about whether and how 
the principle is respected in the electromagnetic radiation from 
a uniformly accelerated classical charge, or for a charge in free 
fall.  (See, for instance, \cite{PV}.)  The issues raised there 
are certainly related to those involving Unruh detectors, but 
they are different in some important respects.  Another type 
of model system, somewhat intermediate between charges and 
detectors, involves quantum field theories with reflecting 
boundaries that are allowed to accelerate \cite{moore}.  
For a massless scalar field in two-dimensional space-time,
moving-mirror problems can be solved in closed form 
(nonperturbatively) and have cast 
great light on the more difficult problems of quantum field 
theory in external gravitational fields.
Mirror models yield local expectation values of the 
stress-energy-momentum tensor, therefore providing more detailed 
information than calculations of transition amplitudes that 
require integration over entire worldlines.
Additionally, lacking charge or 
internal structure, the 
moving mirrors provide a simple testing 
ground for general statements of equivalence.

In Sec.\ \ref{sec:atom} we examine by gedankenexperiments the 
issues raised by absolute and relative acceleration of an atom 
and a resonant cavity, where the acceleration may or may not be 
gravitational.  In Sec.~\ref{sec:mirror} we replace the atom by a 
mirror (in dimension $1+1$) and study the radiation produced, as 
manifested by the covariantly renormalized stress-energy-momentum 
tensor of 
the massless scalar field \cite{FD}, which we review in 
Sec.~\ref{sec:2D}.  The most unexpected 
conclusion --- but one necessary to save the principle of 
equivalence --- is that a stationary mirror radiates from the 
point of view of an accelerated observer, though only if that 
observer's environment is initially some state 
similar to the ``Rindler 
vacuum'' \cite{F73,B75,U76}. (The mirror 
could equally well
be moving at a constant velocity, but since we 
will 
be considering only one inertial mirror, we 
may work in
its inertial frame, where it is \emph{stationary}).  
This effect of the mirror may be regarded as 
reflection of the negative energy density already present in the 
Rindler ground state.  In Sec.~\ref{sec:concl} we observe that in 
essence the effect has 
already been calculated by Davies long ago \cite{D75}, and then 
we draw some conclusions about the equivalence principle.

We use natural units, $c=1=\hbar$.  
For consistency with the original literature, such as \cite{FD},
we use the metric signature in which the minus sign is associated  
with the spatial dimension.

\section{Four experiments with an atom and a cavity} 
\label{sec:atom}

\begin{figure}
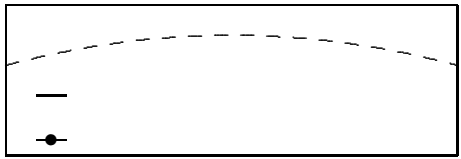

$$\beginpicture
\setcoordinatesystem  units <2cm,2cm> 
\putrule from 0 0 to 0 1
\putrule from 0 0 to 3 0
\putrule from 3 0 to 3 1
\putrule from 0 1 to 3 1
\putrule from .2 .1 to .4 .1
\putrule from .2 .4 to .4 .4
\put{$\bullet$} at .3 .1 
\setquadratic\setdashes
\plot 0 .6
      1.5 .8
      3 .6 /
\endpicture
$$
\caption{An atom is in its ground state, with an unoccupied 
excited state above it. The fundamental mode of the cavity is 
drawn with a dashed curve to indicate that it is unoccupied.}
\label{fig:groundstate}\end{figure}

 Consider a two-level atom in its ground state, inside a 
resonant cavity capable of acting as a photon detector 
(Fig.~\ref{fig:groundstate}). 

\emph{Experiment 1:}
 Suppose that the atom is accelerated; 
is there a chance that the cavity mode will be excited? 

\begin{figure}
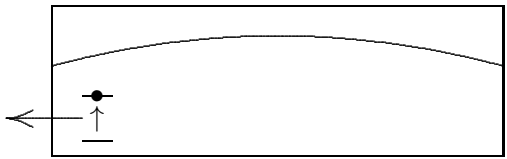

$$\beginpicture
\setcoordinatesystem  units <2cm,2cm> 
\putrule from 0 0 to 0 1
\putrule from 0 0 to 3 0
\putrule from 3 0 to 3 1
\putrule from 0 1 to 3 1
\setquadratic
\plot 0 .6
      1.5 .8
      3 .6 /
\arrow <10pt> [.2,.67] from .2 .25 to -.3 .25
\putrule from .2 .1 to .4 .1
\putrule from .2 .4 to .4 .4
\put{$\bullet$} at .3 .4 
\put{$\uparrow$} at .3 .25
\endpicture
$$
\caption{An accelerated atom excites as an Unruh--DeWitt 
detector \cite{U76,dW79}, 
thereby emitting Unruh--Wald radiation.}
\label{fig:exp1}\end{figure}

Despite some remaining pockets of dissent, the answer is now 
generally agreed to be \emph{yes}\cite{UW}:
  The atom may move to its excited state and emit a photon 
(Fig.~\ref{fig:exp1}).

\emph{Experiment 2:}
 Now suppose that the atom is stationary and the 
cavity is accelerated; can the cavity be excited?

\begin{figure}
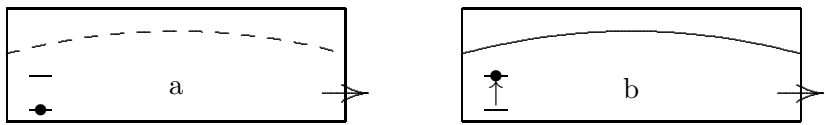

$$\beginpicture
\setcoordinatesystem  units <1.5cm,1.5cm> 
\putrule from 0 0 to 0 1
\putrule from 0 0 to 3 0
\putrule from 3 0 to 3 1
\putrule from 0 1 to 3 1
\arrow <10pt> [.2,.67] from 2.8 .25 to 3.2 .25
\putrule from .2 .1 to .4 .1
\putrule from .2 .4 to .4 .4
 \put{$\bullet$} at .3 .1 
\setquadratic\setdashes
\plot 0 .6
      1.5 .8
      3 .6 /
\put{a} at 1.5 .3
\endpicture
\qquad\qquad
\beginpicture
\setcoordinatesystem  units <1.5cm,1.5cm> 
\putrule from 0 0 to 0 1
\putrule from 0 0 to 3 0
\putrule from 3 0 to 3 1
\putrule from 0 1 to 3 1
\setquadratic
\plot 0 .6
      1.5 .8
      3 .6 /
\arrow <10pt> [.2,.67] from 2.8 .25 to 3.2 .25
\putrule from .2 .1 to .4 .1
\putrule from .2 .4 to .4 .4
\put{$\bullet$} at .3 .4 
\put{$\uparrow$} at .3 .25
\put{b} at 1.5 .3
\endpicture
$$
\caption{Two opinions about thought experiment 2.  
(a) The atom is stationary, so it does not 
excite or radiate.  
(b) The atom is excited and radiates, much as in experiment~1.}
\label{fig:exp2}
\end{figure}

At this point a difference of opinion may emerge. One natural 
answer is, ``Of course not.  Nothing is happening 
to the atom.'' (See Fig.~\ref{fig:exp2}(a).)
In \cite{SFLPSS,SBFP} it is argued that the answer is 
\emph{yes} (Fig.~\ref{fig:exp2}(b)).
But for the moment let's accept the negative answer 
 and  explore its consequences.

Let us now suppose that 
these accelerations are caused by a 
gravitational field (as opposed to, say, a rocket motor).
In other words, the formerly accelerated bodies are 
now considered in free fall, and the formerly
stationary bodies are now 
\emph{supported}  by some force, to avoid falling.

\emph{Experiment 3:}
The atom is falling and the cavity is supported.
Can the cavity  be excited? 

Now the party of \emph{no} in regard to experiment~2 may further 
divide into factions.

\begin{figure}
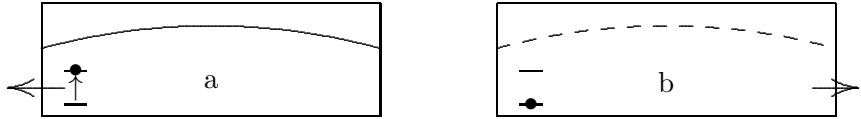

$$\beginpicture
\setcoordinatesystem  units <1.5cm,1.5cm> 
\putrule from 0 0 to 0 1
\putrule from 0 0 to 3 0
\putrule from 3 0 to 3 1
\putrule from 0 1 to 3 1
\setquadratic
\plot 0 .6
      1.5 .8
      3 .6 /
\arrow <10pt> [.2,.67] from .2 .25 to -.3 .25
\putrule from .2 .1 to .4 .1
\putrule from .2 .4 to .4 .4
\put{$\bullet$} at .3 .4 
\put{$\uparrow$} at .3 .25
\put{a} at 1.5 .3
\endpicture
\qquad\qquad
\beginpicture
\setcoordinatesystem  units <1.5cm,1.5cm> 
\putrule from 0 0 to 0 1
\putrule from 0 0 to 3 0
\putrule from 3 0 to 3 1
\putrule from 0 1 to 3 1
\arrow <10pt> [.2,.67] from 2.8 .25 to 3.2 .25
\putrule from .2 .1 to .4 .1
\putrule from .2 .4 to .4 .4
\put{$\bullet$} at .3 .1 
\put{b} at 1.5 .3
\setquadratic\setdashes
\plot 0 .6
      1.5 .8
      3 .6 /
\endpicture
$$
\caption{Two opinions about experiment~3 (among those holding the 
first opinion about experiment~2).  (a) As far as atomic physics 
is concerned, it is the atom that accelerates.
(b) According to general relativity, the atom is in an inertial 
frame and the 
cavity accelerates in the opposite direction.}
\label{fig:exp3}\end{figure}

Answer 1: \emph{Yes}. This is still experiment 1 
(Fig.~\ref{fig:exp3}(a)).  
The radiation by the atom is a matter of electromagnetism and
nonrelativistic quantum mechanics. 
The nature of the applied force doesn't affect that process. 

Answer 2:  \emph{No.}  This is still experiment 2 
(Fig.~\ref{fig:exp3}(b)).  
By the equivalence principle, it is the atom that is at rest 
and the cavity that is accelerating.

\emph{Experiment 4:}
The atom is supported and the cavity is falling.  What happens?

In this case the factions represented in Fig.~\ref{fig:exp3} are 
likely to interchange their positions:  In terms of atomic 
physics, 
the atom is at rest and should not radiate.   In terms of 
relativity, the atom is accelerated and should radiate.  The 
original party of \emph{yes} in Fig.~\ref{fig:exp2} will 
probably have no 
trouble continuing to answer \emph{yes} in all cases.
So there are three theories, offering  \emph{no-yes-no}, 
\emph{no-no-yes}, 
and \emph{yes-yes-yes} as predictions with respect to experiments 
2,~3,~4.

 The last two  schools may have second thoughts when 
they consider that their position compels them to say that an 
atom at rest in a terrestrial laboratory has some probability of 
exciting and radiating.  (In such remarks one is supposed to 
ignore the rotation of the earth.  The issue concerns the radial 
force that keeps the atom from falling through the floor.)
All this sounds  hauntingly familiar to onlookers acquainted with 
the never-ending debates about radiation from a 
 classical accelerating charge, in which case the final paradox 
is usually formulated, ``Is it really true that a charge sitting 
on a table on the earth emits Larmor radiation?''  

Sensing victory, the \emph{no-yes-no} party now issues a 
rebuttal:
``The rest of you are abusing the equivalence principle.
Free fall in a gravitational field 
(metric $ds^2 = z^2\,dt^2 - dz^2$, transverse dimensions ignored)
is not really the same thing as acceleration in flat space-time
(metric $ds^2 = dt^2 - dz^2$).
Don't be misled by popularizations that say, `Special relativity 
shows that velocity is relative, and general relativity shows 
that acceleration is relative.'{''}
The \emph{no-no-yes} party agrees, claiming the equivalence only connects Experiments 1 and 4 and Experiments 2 and 3.
Those critics are right, in that general 
relativity does not relativize 
acceleration in the sense that special relativity
  relativizes velocity.
Experiments 1 and 2 are not 
exactly equivalent.  More quantitatively,
one should observe the behavior of worldlines under the
\emph{Rindler coordinate transformation} \cite{rindler},
\beq t = \rho \sinh \tau, \quad
z = \rho \cosh \tau.  \label{rindtr} \eeq
whose inverse is 
\beq
 \tau = \tanh^{-1} \left(\frac tz  \right)  , \quad
\rho = \sqrt{z^2-t^2}\,    .  \label{rindinv} \eeq
The metric transforms as
\begin{equation}
ds^2 = dt^2 -dz^2 = \rho^2\,d\tau^2 - d\rho^2.
\label{rindmetric}\end{equation}
The worldline $\rho=  1/a$ is the hyperbola
$z^2-t^2= a^{-2}$, corresponding to 
\emph{uniform acceleration}~$a$.
In contrast, the worldline $z= 1/a$ is the same as the curve
\beq \rho = \frac 1{a   \cosh \tau}\,.
\label{daviesrho}\eeq
This formula describes how  a \emph{stationary} body is 
regarded by an accelerated observer.

Nevertheless, we claim that the equivalence principle remains 
\emph{qualitatively} valid in this situation of relative acceleration between an atom and a 
cavity, and that it 
dictates the \emph{yes-yes-yes} conclusions.  
Transition-amplitude calculations 
\cite{SKBFC,BKCFZS,SBFP,LevelPeres1992} back up these 
conclusions. Similar scenarios were discussed by Ginzburg 
and Frolov \cite[Sec.~6]{GF}, with similar conclusions; those 
authors emphasized the importance of being explicit about the 
choice of initial ``vacuum'' state in each scenario.

\emph{Remark:} This is not the same as saying \emph{yes-yes-yes} 
to the corresponding questions 
about classical charges.  That situation turns out to be 
surprisingly muddled, even after the analogous 
quantum-field-theory questions have been settled.  The 
classical charge and 
the Unruh atom differ in two major respects:  The charge has no 
internal degrees of freedom, and the classical 
electromagnetic theory has no distinction between rival 
``vacuum'' states like the one that forms the crux of the Unruh 
theory \cite{U76}.  These differences turn out to make the 
classical-charge problem harder, not easier, to understand.  

In 
the present paper we study primarily an accelerated perfect 
mirror, which has no internal degrees of freedom (by definition 
of ``perfect'') but does have a variety of ``vacuum'' states in 
quantum field theory.  We shall show that the answers to the 
analogous questions are \emph{yes-yes-yes} when the initial 
states of the quantum field are appropriately chosen, thereby 
buttressing and generalizating the conclusions of 
\cite{SBFP,SFLPSS} with a ``simplified'' 
atom (the mirror).

\emph{Remark:} In this section we have avoided the word 
``detector'' as much as possible, because of its ambiguity.
In the original analysis of experiment~2 by Unruh \cite{U76}
the accelerated atom acts as a detector, responding to the quanta 
in the Rindler thermal bath that exists in the usual vacuum state 
of the quantum field (or in the ground state of the cavity, in 
our present scenario).  In \cite{SFLPSS}, on the other hand, the 
cavity is tuned to respond preferentially to a certain mode of 
the field, and hence it can detect quanta emitted by the atom.
Both points of view are correct; they are complementary.  In the 
cases of mirrors and charges, however, only the second 
interpretation makes sense.

\section{The amazing triviality and richness of two-dimensional 
massless quantum field theory} \label{sec:2D}

Massless $(1+1)$-dimensional  models are the source of much of 
what we understand 
about quantum field theory  in curved space or in the presence 
of boundaries or acceleration.
This is so, even though massless 
$(1+1)$-dimensional free
quantum field theory is very special, hence 
potentially misleading if one jumps to general conclusions too 
quickly.
Its special properties make it exactly solvable in most 
circumstances, and that, of course, is the source of its power.

In this paper we are primarily concerned with the effect of 
time-dependent boundaries introduced into flat (Minkowski) 
space-time.  The pioneer paper on this topic is by Moore 
\cite{moore}, who treated two mirror-like points bounding a 
finite spatial interval, thereby predicting the particle creation 
now often called ``dynamical Casimir effect''\negthinspace.
Independently, DeWitt \cite{dW75} described the same effect for a 
single boundary.  This theory was further developed and 
generalized by Fulling and Davies \cite{FD,DF77a}
and applied to a model black hole by Davies, Fulling, and Unruh 
\cite{DFU}.  Davies \cite{D75,DF77b}, in particular, recognized 
that the center of 
coordinates of a $(3+1)$-dimensional spherically symmetric star 
acts mathematically as a Dirichlet boundary that effectively 
accelerates away from outside observers if the star collapses to 
a black hole.
Numerous papers since then have made additional contributions, of 
which 
\cite{walker} (which corrected an error in 
\cite{DF77b}) and \cite{CW,paren,V3,BCG,nicol} are especially 
significant; note also a long series of recent papers by Good et 
al., traceable from the most recent one \cite{GL18}.

The first special property of this theory is that  \emph{every 
two-dimensional manifold is locally conformally flat}:
There are coordinates (nonunique) where the line element takes 
the simple forms
\beq
\aligned ds^2 &= C\,(dt^2-dx^2) \\
&= C(u,v)\, du\,dv, \endaligned
\label{cflat}\eeq
for some function $C$ on space-time,
where the \emph{null coordinates}  are
\beq
u=t-x,\quad v = t+x.
\label{nullc}\eeq
In other words,
\beq
g_{uv} = {\tfrac12}C, \quad g^{uv} = {\tfrac 2C}\,, 
\label{nullm}\eeq
and on-diagonal elements of the metric tensor are zero (thus 
raising or lowering indices converts $u$-components to 
$v$-components and vice versa).
The lines of constant $u$ or $v$ are \emph{light rays}.
Any other such coordinate system must be just a relabeling of 
these rays:
$u=f(u^*),  v= g(v^*)$ leads to 
\beq
\aligned ds^2 
&= f'(u^*) g'(v^*) C\bigl(f(u^*),g(v^*)\bigr)\, 
du^*\, dv^*  \\ 
&= C^*(u^*,v^*)\, du^*\, dv^* \endaligned
\label{conftr}\eeq
for some new conformal factor $C^*$.

The second special property is that  
\emph{the massless wave equation is conformally 
invariant}:
$0=g^{\alpha\beta} \nabla_\alpha \nabla_\beta \phi = 
\frac 4C \, \nabla_u \nabla_v \phi$,
so
$\nabla_u \nabla_v \phi =0$ if and only if $\nabla_{u^*} \nabla_{v^*} \phi 
=0$.
(Note that a Klein--Gordon mass would ruin this property:
$\nabla_{t^*}{}\!^2\phi - \nabla_{x^*}{}\!^2\phi + m^2\phi=0$ 
becomes the nontrivial
$\nabla_{u} \nabla_{v} \phi + C(u,v) m^2 \phi = 0$.)
Furthermore, the normal modes are elementary,  
simply built out of plane waves 
$ e^{-ip_u u}$ and  $e^{-ip_v v}$, or 
 $e^{-i(\omega t - k x)}$ ($\omega=|k|$)
and their negative-norm conjugates.
The wave $e^{-ip_u u} = e^{-ip_u(t-x)}$ is right-moving,
while $e^{-ip_v v} = e^{-ip_v(t+x)}$ is left-moving.
\emph{How} to build normal modes from these pieces depends on 
global geometry and 
boundary conditions.  For a reflecting boundary at $x=0$ one 
needs
 $\phi \propto e^{-i\omega t} \sin(kx)$, for instance.

Additionally,  creation operators of normal modes in 
one frame in general correspond to superpositions of creation and 
annihilation operators in another; for instance 
$e^{i p_{u} u} 
\propto \sum_k\left[ \alpha_{p_{u} k} e^{i k u^*} + 
\beta_{p_{u} k}e^{-i k u^*} \right]$ implies that the annihilation 
operator in the $(u,v)$-basis ($b_{p_{u}}$) can be written in terms 
of annihilation and creation operators in the $(u^*,v^*)$-basis 
($a_k$ and $a_k^\dagger$, respectively):
\begin{equation}
  b_{p_{u}} = \sum_k \left[ \alpha_{p_{u} k} a_k + 
\beta_{p_{u} k}^* a_k^\dagger \right].
\end{equation}
The vacuum annihilated by $a_{k}$ (call it 
$\ket{0_a}$) 
does not have single excitations with respect to $b_{p_{u}}$ 
(whose vacuum is $\ket{0_b}$):
 $\braket{ 0_b | b_{p_{u}} | 0_a} = 0$ 
while $\braket{ 0_a | b_{p_{u}}^\dagger b_{p_{u}} | 0_a} = 
\sum_k|\beta_{p_{u} k}|^2$.
Therefore, we can already say that
 the radiation from a mirror is of a slightly different 
character
from the atomic radiation in Refs.~\cite{SFLPSS,SBFP}.

Please note:  The physics \emph{does} depend on $C$.
The physical world as a whole is not conformally 
invariant.
Measuring instruments know about the metric tensor.
If that were not so, all two-dimensional models would be just 
like static, flat space, and there would be no ``effects''
to be named after Unruh, Moore, and Hawking.

\emph{Example:}  To fit the Rindler coordinate system of 
Eqs.~(\ref{rindtr})--(\ref{rindmetric}) into this framework,
define $\zeta$ by $\rho = e^{\zeta}$.  Then (\ref{rindmetric}) 
takes (in 2-dimensional space-time) the manifestly conformally 
flat form
\beq
ds^2 = dt^2-dz^2 = e^{2\zeta} (d\tau^2-d\zeta^2),
\label{crindmetric}\eeq
and the stationary worldline (\ref{daviesrho}) becomes
\beq
\zeta= -\ln (a   \cosh \tau). 
\label{cdaviesrho}\eeq
(Note that the coordinates $(\tau,\zeta)$ --- which range 
over all of $\mathbb{R}^2$ ---cover only one 
quadrant, $z>|t|$, of the original space.)
Identifying the $x$ of the general formalism \eqref{nullc} with $z$, and the 
starred coordinates \eqref{conftr} there with barred ones here, one sees that
\begin{equation}
  u = - e^{-\bar u}, \quad v = e^{\bar v},
\quad\mbox{where}\quad \bar u = \tau - \zeta, \quad 
\bar v = \tau + \zeta, 
\label{uvbar}\end{equation}
and the new conformal factor is
\beq
\bar C(\bar u,\bar v) = e^{\bar v - \bar u}.
\label{rindconffac}\eeq

What would be observable in our imaginary two-dimensional world?
Presumably not particles, since following the textbook 
prescription for quantizing the field in the $(\tau,\zeta)$ 
coordinates and in the $(t,z)$ coordinates yield different 
results \cite{F73,U76}.  
The Minkowski vacuum is \emph{not} the ground state of the 
Fock space built on Rindler modes, 
$\bar\phi \propto e^{i(-\omega\tau + k\zeta)}$.
It is a mixed state of nonzero temperature proportional to the
acceleration of a trajectory of fixed $\zeta$.
Unruh \cite{U76} observed that there is real physics, not just a 
mathematical artifact, in this nonstandard quantization:
A detector at $\zeta$ observes 
a thermal bath at that temperature.
But for a general  coordinate system the quanta 
have no such clear physical interpretation,
if they can be defined at all,
 and the real problem becomes
 not uniqueness of a natural particle interpretation, but 
existence.
Therefore, the challenge to the subject in the 1970s was to 
define \emph{field observables} that are independent of any 
particle concept and of any coordinate system (except, of course, 
for standard local transformations of tensor components).

Because the primary interest in field theory in curvilinear 
coordinates or curved spaces stems from gravitational physics,
it was natural to concentrate on the 
\emph{stress-energy-momentum tensor}, the source of the 
gravitational field, as the most important observable.
In a standard Cartesian frame in two dimensions,
this tensor has the structure
\[T_{\alpha\beta} = \begin{pmatrix} T_{tt} & T_{tx} \\ T_{xt} & 
T_{xx} 
\end{pmatrix},\]
where $T_t^t$ has the interpretation of energy density,  
$T_x^x$ that of pressure,
and the off-diagonal components those of energy 
flux and  momentum density.
In generic index notation,
\[ T_{\alpha\beta} = \partial_\alpha\phi \partial_\beta\phi 
- \tfrac12 g_{\alpha\beta} 
\,\partial_\lambda\phi \partial^\lambda \phi .\]
In null coordinates, we have
\beq
T_{uu} = \partial_u\phi\,\partial_u \phi, \quad 
T_{vv} = \partial_v\phi\,\partial_v \phi, 
\label{nullstress}\eeq
and, at least classically,
\[T_{uv} = \partial_u\phi\,\partial_v \phi 
-\partial_u\phi\,\partial_v \phi
 =0 . \]
$ T_{uu}$ has the interpretation of \emph{rightward flux},
$T_{vv}$ that of  \emph{leftward flux}.
(Recall that $T^{uu} = \frac4{C^2} T_{vv}$, etc.)

The field equation implies the \emph{conservation law}
$\partial_\alpha T^{\alpha\beta} =0$.
Written out in null coordinates, this is
\beq
\partial_uT_{vv} = -\partial_v T_{uv} - C^{-1} \partial_v C\, 
T_{uv} 
\label{nullcons}\eeq
and the analogous equation for $\partial_vT_{uu}\,$. 
Classically, and in quantum field theory in flat space-time, we 
have $T_{uv} =0$ and hence $T_{vv}$ independent of $u$ and vice 
versa. 
In quantum field theory with spatial curvature, however,
$T_{uv}$ will turn out to be nonzero.

To provide a partial explanation of these claims, we step back a 
bit to the generic conformally flat metric, 
\beq ds^2 = C^*(u^*,v^*)\,du^*\,dv^*\label{genconf}\eeq
(cf.\ (\ref{cflat}) and (\ref{conftr})).
In what follows we shall assume that either $(u^*,v^*)$ range 
over all of $\mathbb{R}^2$, or that
$u^*=t^*-x^*$, $v^* = t^*+x^*$, $-\infty<t^*<\infty$, and
$0<x^*<\infty$ with the boundary condition $\phi(t^*,0)=0$ 
imposed on the quantum field.
In those two situations the  textbook quantization can be 
carried out, using spatial eigenfunctions $e^{ikx^*}$ or
$\sin(kx^*)$, respectively; for details see \cite{FD,DF77a}.
One has then a Fock vacuum relative to the starred coordinate 
system, and one can calculate expectation values in it, such as
$\langle T_{vv}\rangle  =
\langle(\partial_v\phi)^2\rangle$.
As always, divergences must be removed.  
The prescription (inherited from the 1970s) is that this must be 
done \emph{in a covariant manner} that 
reduces to the known right answer in flat, or 
initially flat, space-time and also satisfies the conservation 
law, $\nabla_\alpha T^{\alpha\beta} =0$.
The covariant prescription uses the metric structure of flat 
space, since it requires expanding the two-point function
$\langle \phi(t,x) \phi(t',x')\rangle$ in terms of the geodesic 
separation between the two points;
hence it is \emph{not} conformally invariant.
The conservation requirement resolves some ambiguities in the 
renormalization prescription, and it forces the famous
\emph{trace anomaly}:
\beq
T^\alpha_\alpha = \frac4C\, T_{uv}  = -\,\frac R{24\pi} \,,
\label{tranom}\eeq
where $R$ is the \emph{Ricci curvature scalar},
\beq
R=\frac4{C^3} (C \,\partial_u\partial_v C - \partial_uC\, 
\partial_vC)
= -\,\frac4C \,\partial_u\partial_v (\ln C) = -\Box_g(\ln C).
\label{ricci}\eeq
In view of (\ref{nullcons}),
(\ref{tranom}) says that  $R$ 
acts as a \emph{source} in  first-order differential equations 
satisfied by $T_{uu}$ and $T_{vv}$ (an insight due to Unruh 
\cite{DFU}).  
$R$ itself is not dynamical (for a given geometry), but it 
influences how $T_{\alpha\beta}$ propagates.

The upshot of these calculations \cite{FD,DFU,DF77a} is
\beq\aligned
\langle T_{\alpha\beta}\rangle &= 
\theta_{\alpha\beta} -\frac1{48\pi}\, R\, g_{\alpha\beta}\,, \\
 \theta_{uu} &= \frac1{24\pi C^2} [C \partial_u{}\!^2 C -
\tfrac32(\partial_u C)^2],\\
\theta_{vv}& = \frac1{24\pi C^2} [C \partial_v{}^2 C -
 \tfrac32(\partial _v C)^2] ,\\
\theta_{uv}& = 0. 
\endaligned\label{stresstensor}\eeq
(When two mirrors are present, the spectrum of normal modes 
becomes discrete, but the only effect of that is to add to 
(\ref{stresstensor}) a term representing Casimir stress-energy.
We expect to treat that situation in future work \cite{WSF}.)

In the Rindler example, with conformal factor 
(\ref{rindconffac}), one gets from (\ref{stresstensor}) that
\beq
\langle T_{\bar v \bar v} \rangle = \theta_{\bar v \bar v} = 
-\,\frac 1{48\pi} 
= \langle T_{\bar u \bar u} \rangle \,.
\label{rTnull}\eeq
In a local orthonormal frame aligned with the curvilinear 
coordinates, this is 
\beq
\langle T^{\tau}_{\tau}\rangle = -\,\frac1{24\pi}\, e^{-2\zeta} 
= -\,\frac1{24\pi \rho^2}= - \langle T^{\zeta}_{\zeta}\rangle 
\,.
\label{rT}\eeq
(Recall that $\rho$ is the distance from the horizon at 
$x=|t|$, or $\zeta=-\infty$.)
The stress is traceless and the energy is negative.
This Rindler-space vacuum energy is singular at the horizon.
  In the true Minkowski vacuum state, it is 
cancelled by the positive energy of the Unruh thermal bath.

In summary, for any conformal coordinate system (in a 
two-dimensional space-time) that has the right global properties 
to permit the standard Fock-space construction to be performed,
there results a ``vacuum'' state peculiar to that coordinate 
chart, along with creation operators to generate all the other 
states of definite particle number.  There is a conserved 
renormalized
energy-momentum tensor operator, independent of which conformal 
coordinate system is chosen, whose  expectation value in the Fock 
vacuum of any particular such construction is given by 
(\ref{stresstensor}).  If the space-time is flat ($R=0$), the 
off-diagonal component in the null frame, $T_{uv}\,$, is zero;
more generally, that component is proportional to the trace of 
the tensor, which is independent of the state (a fixed multiple 
of~$R$). 

The role of coordinate systems in this discussion may appear 
suspicious:  does it not fly in the face of the modern 
understanding of general relativity?
No. 
Coordinates
are just a tool to make calculations feasible.
The real point is \emph{which initial state} of the field 
is assumed in a calculation.
It is a special property of the two-dimensional massless field 
that any conformal coordinate system defines a state (in the 
globally hyperbolic space-time region covered by those 
coordinates) that has many of the formal properties of the usual 
vacuum state of a free field.
 Barcel\'o et al.\ \cite{BCG} have shown how the construction can 
be reexpressed in manifestly covariant terms, replacing the 
distinguished coordinate system by a distinguished timelike 
vector field.

\section{Radiation from a stationary mirror} \label{sec:mirror}

With the tools defined in Sec.~\ref{sec:2D}, we can address the 
problem of a stationary mirror ``from the point of view 
of an accelerated observer.'' By this we do not mean just that 
the observer uses the Rindler hyperbolic coordinates [16]; of 
course, in principle a calculation can be done in any coordinate 
system, with equivalent results.  Rather, we mean that the 
initial state of the quantum field is a vacuum associated with 
the timelike Killing vector field of the Rindler system.  Ideally 
this environment would be induced by doing experiments inside a 
cavity with perfectly reflecting walls, which has been uniformly 
accelerated for all time, and has been refrigerated to remove all 
excitations of the scalar field.
A  cavity of length $L$ has a 
discrete set of modes with 
eigenfrequencies $\omega_n \propto n /L$ as measured with the 
Rindler timelike Killing vector $\partial_\tau$; 
all these modes are initially unoccupied. We 
are interested in what occurs when 
the mirror passes through the cavity, not the effects of the 
finite size, so we take $L \gg 1/a$, leaving us with the entire 
Rindler wedge at zeroth order. Therefore, we will focus on how 
the stationary mirror excites the Rindler vacuum in the part of 
the 
wedge to the right of the  mirror. 
%However, this means 
%that our analysis will necessarily break down (1) near the 
%Rindler 
%horizon (the cavity is not on a null trajectory) and (2) on the 
%light ray originating from when the mirror passes into the 
%Rindler wedge (the mirror enters the cavity sometime after 
%that). 
%This breakdown is well controlled, though. In fact, we will 
%shortly discuss singularities and sharp features in these 
%regions; they are present yet rounded off (in a nonuniversal 
%manner) by the finite acceleration and extent of the cavity.

% First, we will recall the   Eqs.~\eqref{uvbar},\eqref{rindconffac}.

Within the Rindler wedge, we assume that the cavity is initially in the Rindler 
vacuum 
defined by the normal modes $e^{-i \omega \bar u}$ and $e^{-i 
\omega \bar v}$ with $(\bar u, \bar v)$ defined in \eqref{uvbar}.

\begin{figure}
\begin{center}
  \begin{tikzpicture}

   \foreach \x/\xname/\tname/\uname/\vname in {0/$z$/$t$/$u$/$v$,%
                                               5/$\zeta$/$ \tau$/$\bar u$/$\bar v$}{%
   \draw[->] ({-1.5+\x},0) -- ({1.5+\x},0) node(xline)[right] {\xname};
   \draw[->] ({0+\x},-1.5) -- ({0+\x},1.5) node(yline)[above] {\tname};
   \draw[color=red,->] ({-1.5+\x},-1.5) -- ({1.5+\x},1.5) node [anchor=south west] {\vname};
   \draw[color=red,->] ({1.5+\x},-1.5) -- ({-1.5+\x},1.5) node [anchor=south east] {\uname};}

   \draw[color=violet,thick] (0.5,-1.5) node[below] {Mirror} -- (0.5,1.5);
   \draw[domain=-0.9:0.9,variable=\x, smooth, violet, thick] plot ({5+0.5*ln(1+\x)+0.5*ln(1-\x)},{-0.5*ln(1+\x)+0.5*ln(1-\x)}) node [below]{Mirror};

   \draw[dashed] (0,0.5) node [left] {$\tfrac 1a$}  -- (0.5,0.5);
   \draw[dashed] (0,-0.5) node [left] {$-\tfrac 1a$} -- (0.5,-0.5);

   \draw[dashed] ({5-1.5+ln(2)},-1.5) -- (6.5,{1.5-ln(2)}) node [right] {$\bar u = \log(a/2)$};
   \draw[dashed] ({5-1.5+ln(2)},1.5) -- (6.5,{-1.5+ln(2)}) node [right] {$\bar v = -\log(a/2)$};

  \end{tikzpicture}
  \end{center}
  \caption{A stationary mirror at $z = 1/a$ in Minkowski coordinates
 (left) enters the right Rindler wedge for $-1/a < t < 1/a$. 
In this interval, the trajectory of the mirror in Rindler coordinates 
is pictured (right). \label{fig:rindlerPlate}}
\end{figure}
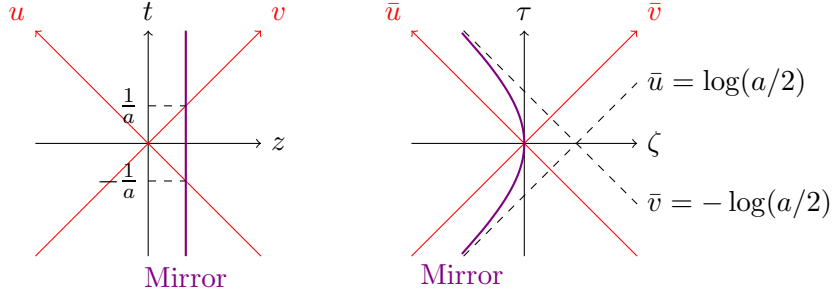
As for the mirror's trajectory, it enters and leaves the Rindler 
wedge as seen in Fig.~\ref{fig:rindlerPlate}. In flat space, the 
trajectory is characterized by its constant spatial coordinate 
$z=1/a$, or by its path in (Minkowski) null coordinates 
$[U(t),V(t)] = (t - 1/a, t+1/a)$. Transforming to Rindler 
coordinates using \eqref{uvbar}, we can find the path in Rindler 
null coordinates:
\begin{equation}
 \begin{split}
  \bar U(t) & = \log a - \log(1 - a t), \\
  \bar V(t) & = \log(1 + a t) - \log a.
 \end{split}
\label{ndaviesrho}\end{equation}
(We use capital letters for the functions that define a 
particular trajectory, and the corresponding lower-case letters 
for the associated coordinates in a chart.
Formulas (\ref{ndaviesrho}) are equivalent to (\ref{daviesrho}) 
or (\ref{cdaviesrho}).)

At $t = -1/a$, the mirror enters the Rindler wedge. At this 
point, notice that $\bar U[(-1/a)^+] = \log(a/2)$ while $\bar 
V[(-1/a)^+] \rightarrow - \infty$. Similarly, when the mirror 
exits the Rindler wedge at $t = 1/a$, $\bar V[(1/a)^-] = 
-\log(a/2)$ and $\bar U[(1/a)^-] = \infty$. These indicate past 
and future null asymptotes as pictured in 
Fig.~\ref{fig:rindlerPlate}.

Now, we can use conformal transformations of the form mentioned 
in \eqref{conftr} along with the remarkable property that the 
massless wave equation is conformally invariant to find a new set 
of coordinates $(\hat u, \hat v)$ with the following property: 
the mirror is at the \emph{constant} coordinate position $\hat x 
= 0$ defining time and space coordinates $(\hat t, \hat x)$ by 
$\hat u = \hat t - \hat x$ and $\hat v = \hat t + \hat x$. Once 
this coordinate transformation is found, the normal modes are 
simply $\phi \propto e^{-i\omega \hat t}\sin(k \hat x)$, and the 
analysis leading to \eqref{stresstensor} applies.

There is in fact a continuum of such coordinate transformations 
$(\bar u, \bar v) \mapsto (\hat u, \hat v) = (p(\bar u), q(\bar v))$ 
(one of which returns us to Minkowski coordinates), and they all must satisfy
\begin{equation}
  p[\bar U(t)] - q[\bar V(t)] = 0, \label{eq:constcoord}
\end{equation}
which is just a restatement of $\hat x = 0$.
For the accelerated cavity defined with the right-moving normal 
modes $e^{-i \omega \bar u}$ and left-moving normal modes $e^{-i 
\omega \bar v}$, the left-movers should remain equally spaced and 
unaffected while the right-movers will be distorted by the mirror 
for $u>\log(a/2)$ (they have been reflected off of the mirror). 
Physically then, we expect $q(\bar v) = \bar v$ and 
\eqref{eq:constcoord} can subsequently be solved such that 
formally
\begin{equation}
  p = \bar V \circ \bar U^{-1},
\end{equation}
and for our particular case,
\begin{equation}
  p(\bar u) = \log(2 - a e^{-\bar u}) - \log(a), \quad \bar u > \log(a/2) 
\label{eq:p-traject}
\end{equation}
% In order to find what radiation a stationary mirror might give off, 
% we make use of null coordinates.
% Beginning from Minkowski coordinates, we define the \emph{null coordinates} as
% \begin{equation}
%   u = t - z, \quad v = t - z.
% \end{equation}
% Then the transformation to Rindler coordinates takes the simple form
% \begin{equation}
%   u = - e^{-\bar u}, \quad v = e^{\bar v}.
% \end{equation}
% The metric can be written in these coordinates as
% \begin{equation}
%   ds^2 = du\, dv = e^{\bar v - \bar u} d\bar u \, d\bar v,
% \end{equation}
% where we define $\bar C(\bar u,\bar v) = e^{\bar v - \bar u}$ as the 
% \emph{conformal factor}.
% A stationary plate at $z = 1/a$ in null coordinates follows the path 
% $[U(t),V(t)] = (t - 1/a, t + 1/a)$ in Minkowski coordinates.
% On the other hand, in Rindler coordinates $[\bar U(t), \bar V(t)]$ where
% \begin{equation}
%  \begin{split}
%   \bar U(t) & = \log a - \log(1 - a t), \\
%   \bar V(t) & = \log(1 + a t) - \log a.
%  \end{split}
% \end{equation}
% Reparametrizing the curve by its $\bar u$-coordinate in Rindler coordinates, 
% we obtain the parametrization $[\bar u, p(\bar u)]$ with 
% $p = \bar V \circ \bar U^{-1}$.
% Working out what $p$ is leads us to
% \begin{equation}
%   p(\bar u) = \log[2 - a e^{-\bar u}] - \log a, \quad \bar u > \log(a/2), 
% \label{eq:p-traject}
% \end{equation}
% which asymptotes in the past to the null trajectory $\bar u = \log(a/2)$ 
% in Rindler coordinates.

To determine the radiation leaving the mirror to the right 
($z>1/a$), we calculate the expectation value of the energy 
momentum tensor $\langle T_{\mu\nu} \rangle$ with \eqref{stresstensor}. 
As previously discussed, this requires the conformal factor in the hatted
coordinates,
\begin{equation}
  ds^2 = \hat C(\hat u,\hat v)\, d\hat u\, d\hat v = f'(\hat u) 
\bar C(f(\hat u), \hat v)\, d\hat u \, d \hat v,
\end{equation}
where $f = p^{-1}$.
If we define the functional
\begin{equation}
   F_x(f) = \frac{\partial_x^2 f(x)}{f(x)} 
- \frac32\left(\frac{\partial_x f(x)}{f(x)} \right)^2,
\end{equation}
then we have
\begin{equation}
  \langle T_{\hat u \hat u} \rangle = \frac1{24\pi} 
F_{\hat u}[\hat C], \quad \langle T_{\hat v \hat v} \rangle 
= \frac1{24\pi} F_{\hat v}[\hat C], \quad
    \langle T_{\hat u \hat v} \rangle = \langle T_{\hat v \hat u} \rangle = 0.
\end{equation}
It can then be checked that 
\begin{equation}
   F_{\hat u}(\hat C) = f'(\hat u)^2 F_{f(\hat u)}(\bar C)
  + F_{\hat u}(f').
\end{equation}
Additionally, if $p$ is the inverse of $f$, then
$F_{\hat u}(f') = - f'(\hat u)^2 F_{f(\hat u)}[p']$.
Written in terms of the $\bar u$, this implies
\begin{equation}
    F_{\hat u}(\hat C) = \frac1{p'(\bar u)^2} [ F_{\bar u}(\bar C)
 - F_{\bar u}(p')].
\end{equation}

The stress-energy tensor in the original Rindler coordinates can 
be found by using the coordinate transformation 
$\langle T_{\bar u \bar u} \rangle = p'(\bar u)^2 \langle T_{\hat 
u \hat u} \rangle$.
Therefore,
\begin{equation}
  \langle T_{\bar u \bar u} \rangle = -\, \frac1{48\pi} 
\frac{a^2 e^{-2 \bar u}}{(2 - a e^{-\bar u})^2}\,, \quad  
\langle T_{\bar v \bar v}\rangle = -\,\frac1{48 \pi}\,, \quad 
\bar 
u> \log (a/2).
\label{ourTRind}\end{equation}
A further simple transformation
takes us back to (null) Minkowski coordinates,
 where we find
\begin{equation}
\begin{split}
  \langle T_{uu} \rangle &= -\,\frac1{48 \pi} \frac{a^2}{(2 + a 
u)^2} \quad\text{for }
 u> -\, \frac2 a \,,\\
  \langle T_{vv} \rangle &= -\,\frac1{48 \pi} \frac1{v^2}
\quad\text{for } v> 0.
\end{split}
\label{ourTmink}\end{equation}

\begin{figure}
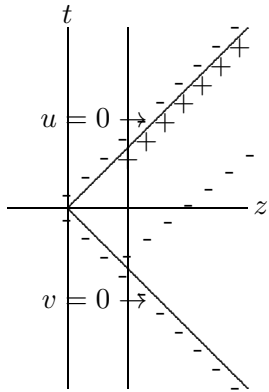

$$\beginpicture
\setcoordinatesystem  units <.8cm,.8cm> 
\putrule from -1 0 to 3 0
\putrule from 0 -3 to 0 3
\putrule from 1 -3 to 1 3
\plot 0 0
         3 3 /
     \plot 0 0
        3 -3 /
    \setplotsymbol({+})  \setdots <10pt>
        \plot 1 .8
                3 2.8 /
       \setplotsymbol({-})  \setdots <10pt>
        \plot 1 -.8
                3.2 1 /
\plot 0 .2
      3 3.2 /
\plot 0 -.2
      3 -3.2 /
\put{$u=0 \rightarrow$} [r] <-4pt,0pt> at 1.5 1.5
\put{$v=0  \rightarrow$} [r] <-4pt,0pt> at 1.5 -1.5
\put{$z$} [l] <2pt, 0pt> at 3 0
\put{$t$} [b] <0pt,2pt> at 0 3
\endpicture$$
\caption{In the absence of the mirror, the Rindler vacuum 
possesses a negative vacuum stress (\ref{rTnull}), (\ref{rT}) 
concentrated near the horizon (here indicated outside the horizon 
for visual clarity, but actually inside).  When the mirror is 
introduced at $z=1/a$ and the region to its right is initially in 
a Rindler vacuum, an outgoing burst of positive stress cancels 
the Rindler stress along the future horizon, while the incoming 
stress along the past horizon is reflected into the region from 
the point where the mirror enters the Rindler wedge.}
\label{fig:Rindrad}\end{figure}

In these formulas, notice first the (negative) flux moving to 
the left, $\langle T_{\bar v\bar v}\rangle$ or $\langle 
T_{vv}\rangle$;
it is well known for the Rindler vacuum in the absence of a 
mirror. In that case there would be a similar flux to the right,
$\langle T_{uu}\rangle = -\,\frac1{48 \pi} \frac1{u^2} $,
which is divergent at $u=0$.
This term has been removed by the presence of the mirror, but it 
also 
seems to have been reproduced at an earlier retarded time,
$u = -  2/ a$. 
A physical explanation of this early negative-energy burst is 
that  the divergent incoming flux 
$\langle T_{vv} \rangle$ has reflected off  the mirror at 
$z=1/a$.
On the other hand, the cancellation of the  
divergence at $u = 0$, present in the normal Rindler vacuum, 
might have been predicted because of energy conservation (for a 
mirror stationary in Minkowski space) and the absence of any incoming 
flux at $v=2/a$ to be reflected.
See Fig.~\ref{fig:Rindrad}.

Any realistic experiment is limited not only in space (by cavity 
walls) but also in time.  The Rindler horizons and the 
accompanying singularities, consequently, are not actually inside 
the experimental region.  

\section{Discussion}\label{sec:concl}

\subsection{The work of Davies} 

The 1975 paper of Davies \cite{D75} has usually been regarded as 
a precursor of the 1976 paper of Unruh \cite{U76}, in which the 
Minkowski-to-Rindler Bogolubov transformation was derived and was 
somehow attributed to the presence of a stationary reflecting 
boundary.  Close examination, however, shows that \cite{D75} is 
actually doing something quite different from \cite{U76} (or 
\cite{F73}, usually cited in the same context) and closer to the 
present paper.  

Davies started from the observation (developed further in 
\cite{DF77b}) that the origin of a spherical coordinate system in 
dimension $(3+1)$ acts after separation of variables as a mirror 
boundary in dimension $(1+1)$, and that in a collapsing black 
hole that boundary ``accelerates'' in Schwarzschild coordinates.
He then followed the black-hole calculation of Hawking \cite{haw}
step by step to show that a stationary mirror in Minkowski space 
radiates as seen by a Rindler observer.  This calculation is 
based on a Bogolubov transformation.  In Rindler coordinates the 
trajectory of the mirror is, of course, the one described by 
(\ref{daviesrho}), (\ref{cdaviesrho}), or (\ref{ndaviesrho}).

In \cite{DF77b} the problem 
was revisited on the basis of the intervening development 
\cite{FD,DFU,DF77a} of a 
physical interpretation of the theory in terms of expectation 
values of the renormalized stress tensor.
As reviewed in  Sec.~\ref{sec:2D}, this approach usually avoids 
explicit Bogolubov transformations.  The paper \cite{DF77b} 
contains the statement, ``[I]t is now clear that the static 
mirror in Rindler space [in \cite{D75}] does not actually produce 
energy,''  and for that reason attention was shifted to the same 
trajectory (\ref{cdaviesrho}) but regarded as embedded in the 
Minkowski background metric ($\eta\to z$ and $\tau\to t$); in 
that case it 
was shown that the trajectory did produce radiation.

In view of the present paper, however, it is now clear that the 
retreat from \cite{D75} was unwarranted and was based on a 
failure to distinguish between Minkowski and Rindler vacuums as 
initial state.  Both Minkowski and Rindler space are flat, so the 
conservation law (\ref{nullcons}) applies with vanishing 
right-hand side (because of (\ref{tranom})).  Therefore, any 
outgoing radiation at future null infinity can be propagated back 
to the mirror (unlike in the case of the model black hole in 
\cite{DFU}); the only difference between the emissions of mirror 
(\ref{cdaviesrho}) in Minkowski space and in Rindler space is  
geometrical factors related to the introduction of local 
orthonormal frames, and it is not possible for the ``radiation'' 
to vanish in one problem and not the other.  The resolution of 
this apparent paradox is that the Rindler-space calculation
(whether by the method of \cite{D75} or that of \cite{DF77b} and 
our Sec.~\ref{sec:mirror})
tacitly starts from \emph{the Rindler vacuum as initial state}.
Thus Davies's Hawking-style calculation revealed real radiation 
from the stationary mirror, embodied in Bogolubov transformation 
coefficients that turned out to be  identical with those of 
\cite{F73,U76}.  What he could not have known at the time is that 
the Rindler vacuum contains a negative stress that (near $u=0$) 
precisely cancels the stress tensor of his radiation.
(In both \cite{D75} and \cite{DF77b} attention was focused on the 
far future, where the relevance to the black-hole situation is 
greatest.  Therefore, the emission at $u\approx -2/a$ in 
(\ref{ourTmink}) was not noted.)

\subsection{The status of the equivalence principle}

The mirror scenario in Sec.\ \ref{sec:mirror} is clearly 
analogous to ``experiment~2'' in Sec.~\ref{sec:atom}.
The role of the atom is played by the stationary mirror that 
enters and exits the Rindler wedge.  An accelerating optical 
cavity within that wedge carries with it its own vacuum, and our 
calculations indicate that in that case, excitations as measured 
by the stress-energy tensor are created; 
cavity modes are excited!
From the perspective of the cavity, the mirror enters violently, 
modifying $\langle T_{\mu\nu}\rangle$.
On the other hand, a stationary observer will  see from the 
outset that the cavity begins with some $\langle T_{\mu\nu}\rangle$
that is reflected off the mirror.  
These conclusions are analogous to the \emph{yes}-\emph{yes}-\emph{yes}
conclusions for the atomic gedankenexperiments.

One already knows that an accelerating mirror radiates into 
a stationary vacuum \cite{FD,DF77b}.  Our converse result thus 
shows that this remains true for an accelerated (Rindler) vacuum 
and stationary mirror.
  Insofar as this principle extends to \emph{all} 
physical phenomena, it resolves the conundrum of which bodies to 
call ``accelerated'' in gravitational problems: those which 
accelerate 
relative to the relevant vacuum.  Its primary 
practical implication is that a uniformly accelerated frame can 
legitimately be treated as a ``static'' frame set in a 
gravitational field; it is this fact that is properly called the 
\emph{qualitative equivalence principle}.  This conclusion is of 
great importance because near a black hole (or in any nontrivial 
static gravitational field) the accelerated frame is the 
physically simpler one, as compared to a frame in free fall 
(which does not experience static conditions).

    Our simple, nonperturbative result verifies that phenomena 
analogous to the radiation by unaccelerated atoms deduced in 
\cite{SFLPSS} and \cite{SBFP} can indeed happen.  If one assumes 
the qualitative 
equivalence principle in generality, such atomic radiation is 
unsurprising, indeed necessary for consistency 
(``\emph{yes-yes-yes}'').  
On the other hand, if one already accepts the logic of 
\cite{SFLPSS} and \cite{SBFP}, 
then the analogous study in the present paper adds to the 
evidence that the principle holds in generality.

Two limitations on this picture
 are important to note.
First, the situations are not \emph{quantitatively} symmetric.
For example, a stationary mirror in an accelerated cavity
is not mathematically the same thing as an accelerated mirror in 
a stationary cavity, as has been understood for a long time 
\cite{FD,DF77b}.
Second, both mirror and atom \cite{SFLPSS,SBFP} studies,
and the general detector analyses of \cite{GF}, show that 
the stationary object does not radiate (in the precise way 
described by the equivalence principle) to the accelerated 
observer or detector unless the latter is initially in its 
Rindler vacuum, or as close to a Rindler vacuum as the modified 
conditions (e.g., cavity walls) permit.
How such a state could be realized in practice is a difficult 
separate question.

\section*{Acknowledgments}

JHW is grateful for the support of the Air Force Office 
for Scientific Research.
This project was stimulated and facilitated by intensive 
discussions among SAF and the research group of Marlan Scully at 
IQSE--TAMU, the research group of George Matsas in Brazil, and
Don Page and Bill Unruh.
IQSE funding aided JHW to attend a workshop in College Station in 
October, 2017, and SAF to attend the PQE conference in Snowbird 
in January, 2018.
Dr.~Unruh observed that the initial burst at $u=-2/a$ is 
physically understandable as specular reflection of the incoming 
Rindler vacuum flux at $v=0$.
Dr.\ Page made helpful comments on the manuscript.


\begin{thebibliography}{99} \frenchspacing


\bibitem{SFLPSS} 
M. O. Scully, S. Fulling, D. Lee, D. Page, W. Schleich, 
and A. Svidzinsky, Quantum optics approach to radiation 
from atoms falling into a black hole, \emph{Proc. Natl. Acad. Sci.} 
{\bf115} (2018) 8131--8136.

\bibitem{haw} S. W. Hawking, Particle creation by black holes,
\emph{Commun. Math. Phys.} {\bf 43} (1975) 199--220.  

\bibitem{SKBFC} M. O. Scully, V. V. Kocharovsky, A Belyanin,
E. Fry, and F. Capasso,
Enhancing acceleration radiation from ground-state atoms via 
cavity quantum electrodynamics,
\emph{Phys. Rev. Lett.} {\bf91} (2003) 243004.

\bibitem{BKCFZS} A. Belyanin, V. V. Kocharovsky, F. Capasso, E. 
Fry, M. S. Zubairy, and M. O. Scully,
Quantum electrodynamics of accelerated atoms in free space and 
in cavities,
 \emph{Phys. Rev. A} {\bf74} (2006) 023807.

\bibitem{U76} W. G. Unruh, Notes on black-hole evaporation, 
\emph{Phys. Rev. D} {\bf14} (1976) 870--892.

\bibitem{UW} W. G. Unruh and R. M. Wald, 
What happens when an accelerating observer detects a Rindler 
particle,
\emph{Phys. Rev. D} {\bf29} (1984) 1047--1056.

\bibitem{SBFP} A. Svidzinsky, J. Ben-Benjamin, S. A. Fulling, and
D. N. Page, Excitation of an Atom by a Uniformly Accelerated 
Mirror through Virtual Transitions, \emph{Phys. Rev. Lett.} 
{\bf121} (2018) 071301.

\bibitem{schutz} B. Schutz, \emph{A First Course in General 
Relativity}, 2nd ed., Cambridge U. Press, Cambridge, 2009.

\bibitem{PV} M. Pauri and M. Vallisneri,
Classical roots of the Unruh and Hawking effects,
\emph{Found. Phys.} {\bf29} (1999) 1499--1520.

\bibitem{moore} G. Moore, Quantum theory of the electromagnetic 
field in a variable-length one-dimensional cavity, \emph{J. Math. 
Phys.} {\bf11} (1970) 2679--2691.

\bibitem{FD} S. A. Fulling and P. C. W. Davies. Radiation from a 
moving mirror in two dimensional space-time:  conformal anomaly, 
\emph{Proc. Roy. Soc. Lond. A} {\bf348} (1976) 393--414.

\bibitem{F73} S. A. Fulling, Nonuniqueness of canonical field 
quantization in Riemannian space-time, \emph{Phys. Rev. D} {\bf7} 
(1973) 2850--2862.

\bibitem{B75} D. G.  Boulware, 
Quantum field theory in Schwarzschild and Rindler spaces,
\emph{Phys. Rev. D} {\bf11} (1975) 1404--1423.

\bibitem{D75} P. C. W. Davies, Scalar particle production in 
Schwarzschild and Rindler metrics, \emph{J. Phys. A} {\bf8} 
(1975) 609--616.

\bibitem{dW79} B. S. DeWitt, Quantum gravity:  the new 
synthesis,  in \emph{General Relativity:  An Einstein Centenary 
Survey}, S. W. Hawking and W. Israel, eds., 
Cambridge U. Press, Cambridge, 1979, pp. 680--745.

\bibitem{rindler} W. Rindler. Kruskal space and the uniformly
accelerated frame, \emph{Am. J. Phys.} {\bf34} (1966) 1174--1178.  

\bibitem{LevelPeres1992} O. Levin, Y. Peleg, and A. Peres, 
Quantum detector in an accelerated cavity, \emph{J. Phys. A} {\bf25} 
(1992) 6471--6481.

\bibitem{GF} V. L. Ginzburg and V. P. Frolov, Vacuum in a
homogeneous gravitational field and excitation of a uniformly 
accelerated detector, \emph{Sov. Phys. Usp.} {\bf30} (1987) 
1073--1095 [\emph{Usp. Phys. Nauk} {\bf153} (1987) 633--674].

\bibitem{dW75} B. S. DeWitt, Quantum field theory in curved 
spacetime,
\emph{Phys. Reports} {\bf 19} (1975) 295--357.

\bibitem{DF77a} P. C. W.  Davies and S. A. Fulling, 
Quantum vacuum energy in two dimensional space-times,
\emph{Proc. Roy. Soc. Lond. A} {\bf354} (1977) 
59--77.

\bibitem{DFU} P. C. W. Davies, S. A. Fulling, and W. G. Unruh,
Energy-momentum tensor near an evaporating black hole,
\emph{Phys. Rev. D} {\bf13} (1976) 2720--2723.

\bibitem{DF77b} P. C. W. Davies and S. A. Fulling,
Radiation from moving mirrors and from black holes,
\emph{Proc. Roy. Soc. Lond. A} {\bf356} (1977) 237--257.

\bibitem{walker} W. R. Walker,
Particle and energy creation by moving mirrors,
\emph{Phys. Rev. D} {\bf31} (1985) 767--774.


\bibitem{CW} R. D. Carlitz and R. S. Willey,
Reflections on moving mirrors, \emph{Phys. Rev. D} {\bf87} (1987)
2327--2335.

\bibitem{paren} R. Parentani, The energy-momentum tensor in 
Fulling--Rindler vacuum,
\emph{Class. Quantum Grav.} {\bf10} (1993) 1409--1415.

\bibitem{V3} M. Visser, Gravitational vacuum polarization. III,
\emph{Phys. Rev. D} {\bf54} (1996) 5123--5128. 


\bibitem{BCG} C. Barcel\'o, R. Carballo, and L. J. Garay, 
Two formalisms, one renormalized stress-energy tensor,
\emph{Phys. Rev. D} {\bf12} (2012) 084001.

\bibitem{nicol} N. Nicolaevici, Unruh effect without Rindler 
horizon,
\emph{Class. Quantum Grav.} {\bf32} (2015) 045013.

\bibitem{GL18} M. R. R. Good and E. V. Linder, 
Eternal  and evanescent black holes and accelerating mirror 
analogs, 
\emph{Phys. Rev. D} {\bf97} (2018) 065006.

\bibitem{WSF}  J. H. Wilson, F. Sorge, and S. A. Fulling,
Tidal and nonequilibrium Casimir effects in a falling Casimir 
apparatus, to appear.


\end{thebibliography}
\end{document}